\documentclass[preprintnumbers,amsmath,amssymb,twocolumn,aps,pre]{revtex4}
\def\referencias{/Users/raul/Library/texmf/myreferences}
 \usepackage{natbib}

\usepackage{graphicx}

\newcommand{\ensemblemean}[1]{\langle\hskip-2pt\langle{#1}\rangle\hskip-2pt\rangle}
\begin{document}

\title{Order parameter expansion study of synchronous firing induced by quenched noise in the active rotator model}

\author{Niko Komin}
\author{Ra\'ul Toral}

\affiliation{IFISC (Instituto de F{\'\i}sica Interdisciplinar y Sistemas Complejos), UIB-CSIC, Campus UIB, 07122 Palma de Mallorca,Spain}
\date{\today}


\begin{abstract}
We use a recently developed order parameter expansion method to study the transition to synchronous firing occuring in a system of  coupled active rotators under the exclusive presence of quenched noise. The method predicts correctly the existence of a transition from a rest state to a regime of synchronous firing and another transition out of it as the intensity of the quenched noise increases and leads to analytical expressions for the critical noise intensities in the large coupling regime. It also predicts the order of the transitions for different probability distribution functions of the quenched variables. We use numerical simulations and finite size scaling theory to estimate the critical exponents of the transitions and found values which are consistent with those reported in other scalar systems in the exclusive presence of additive static disorder.

\end{abstract}

\maketitle

\section{Introduction}
In some cases, a dynamical system with many variables depends on a set of parameters which, although fixed in time, are randomly distributed according to a given probability distribution. The outcome of the system, although deterministic, depends on the actual realization of the set of parameters. The influence of this so--called, depending on the context: quenched noise, static disorder, heterogeneity, variability, diversity, impurities, etc. has been the subject of many investigations. In the last years, some emphasis has been put in identifying those situations in which the presence of the quenched noise induces some sort of macroscopic ordering, such as phase transitions~\cite{BP:01}, patterns~\cite{BL2003}; improves the global response to an external forcing~\cite{tessone2006} or enhances synchrony of firing units~\cite{tessone2007a}. 

Due to the complexity of the problem, the analytical treatments are usually very difficult to be carried out in full detail and most results rely on extensive numerical simulations. However, a recently introduced technique named ``order  parameter expansion'' \cite{deMonteOvidio2002,deMonteOvidioMosekilde2003,deMonteOvidioChateMosekilde2004,deMonteOvidioChateMosekilde2005,iacyel2006,KT2010}
offers a simple approximate way of analyzing the effect of the random quenched terms in the dynamical equations. The approximation scheme allows the reduction of the large number of coupled differential equations for the microscopic variables to just a few effective equations for the relevant macroscopic dynamical variables: the mean values and dispersions from the mean.

It is the purpose of this paper to apply this technique to the study of the active rotator model~\cite{Kuramoto1975,Strogatz2000,RevModPhys.77.137} under the influence of static disorder in the natural frequencies. Previous work~\cite{SK:1986,KS:1995,tessone2007a,ToralTessoneLopes2007} has shown the paradoxical result that intermediate levels of disorder at the microscopic level induce macroscopic order which manifests itself in a coherent firing of the units. Although very sophisticated treatments of this model do exist leading to analytical solutions in some particular cases~\cite{OttAntonsen:2008,OttAntonsen:2009,LafuerzaColetToral2010} (and we will refer to them later in the paper) a particular simple analysis was developed in~\cite{tessone2007a}, where the authors used an expansion of the dynamical equations of the model up to first order in the deviation of the quenched variables from their mean value and identified a self-consistency equation which had to be solved numerically. The order parameter expansion used in the present article expands consistently this analysis up to terms of second order, thus reaching a higher accuracy. The resulting closed system of three differential equations reproduces the ordering abilities of quenched noise in this system and, furthermore, predicts a sharp transition back into the disordered state where no macroscopic order is observed. 

The article is organized as follows. First, in section~\ref{sec:model}, we will define the active rotator model and summarize its main properties. Macroscopic observables that describe the collective behavior are introduced. Then, in section~\ref{sec:method}, the approximation method is applied and conclusions about steady states are drawn. In section~\ref{sec:numSim} we present numerical results that support the previous findings and use the theory of finite--size scaling to determine some of the critical exponents characterizing the transitions. The paper closes with concluding remarks in section~\ref{sec:conclusions}.

\section{Model}
\label{sec:model}

Let us consider a system of globally-coupled active rotators~\cite{Kuramoto1975}, defined by a set of angular variables $\phi_i(t), \,i=1,\dots,N$ which evolve according to the dynamical equations:
\begin{equation}
\label{eq:ActiveRots}
 \dot \phi_i(t)=\omega_i-\sin(\phi_i(t))+\frac{C}{N}\sum_{j=1}^N \sin\left(\phi_j(t)-\phi _i(t)\right) \, .
\end{equation}
$C$ is the coupling constant. The so-called natural frequencies $\omega_i$ are quenched noise, i.e. random variables independently drawn from a probability distribution $g(\omega)$ with mean $\left\langle\omega\right\rangle$. The variance of the distribution, $\sigma^2$, is a measure of the dispersion of natural frequencies amongst the oscillators and measures the degree of intrinsic disorder. We will refer to $\sigma$ as the ``diversity". 

For uncoupled units, $C=0$, a value of $\left|\omega_i\right|>1$ results in a rotating behavior for $\phi_i(t)$. The actual period of rotation is $\frac{2\pi}{\sqrt{w_i^2-1}}$ and the direction of rotation depends on the sign of $\omega_i$: clockwise if $\omega_i<0$ and anti-clockwise otherwise. If $\left|\omega_i\right|<1$, then unit $i$ is in an excitatory regime. In this case there are two fixed points (one stable and the other unstable) located at the two solutions of $\phi_i^{*}=\arcsin(\omega_i)$. When a perturbation is such that it makes variable $\phi_i$ to cross over the unstable fixed point, the subsequent dynamics returns to rest again in the stable fixed point through a full turn of $\phi_i$ on the unit circle (a ``spike'' or a ``pulse''). This is the typical behavior of an excitable system~\cite{Lindner2004}.

When the coupling is active, $C>0$, the dynamics of each unit is influenced by the others which act, effectively, as a perturbation. As a result, individual spikes can be generated. Those spikes can be independent of each other or, alternatively, the units might spike with some degree of synchrony. It is of interest to characterize the global behavior of the system in order to identify the region of parameter space for which synchronized spiking occurs. To this end one usually defines a complex variable which represents the center of mass of all rotators~\cite{Kuramoto1984}:
\begin{equation}
\label{eq:centerMass}
 \rho(t){\rm e}^{\imath\Psi(t)}=\frac{1}{N}\sum_{j=1}^N\rm{e}^{\imath\phi_j(t)}\equiv\left\langle \rm{e}^{\imath\phi_j(t)}\right\rangle.
\end{equation}
Henceforth, $\left\langle\cdots\right\rangle$ denotes an average over the $N$ units. The Kuramoto order parameter $\rho=\overline {\rho(t)}$, where the overline denotes an average with respect to time, differentiates between fully synchronized ($\rho=1$, i.e. $\phi_i(t)=\phi_j(t), \forall i,j$) and desynchronized oscillators ($\rho\approx 0$). When $\rho$ is close to $1$, one still needs to distinguish the rest state where all oscillators are equally constant in time from the coherent firing regime where the units are oscillating synchronously. Amongst other possible measures, one can use the order parameter introduced by Shinomoto and Kuramoto~\cite{SK:1986} as $\zeta=\overline{ \left|\rho(t)e^{i\Psi(t)}-\overline{\rho(t)e^{i\Psi(t)}} \right|}$, which is different from zero only in the case of synchronous firing. Alternatively, and this is the approach followed in this paper, one can measure the average angular speed of the time evolution of the global phase $\Psi(t)$. In the rest state, $\Psi(t)$ is time independent and the angular speed is zero, whereas in the coherent firing regime, $\Psi(t)$ increases with time and the angular speed adopts a non-zero value.

It has been shown that the system of coupled active rotators displays a disorder-induced transition from the global rest state to synchronized firing~\cite{SK:1986,KS:1995,tessone2007a}. Higher levels of disorder lead the system again into unsynchronized firing. The disorder can be originated by the existence of diversity amongst the natural frequencies~\cite{ToralTessoneLopes2007} (as it is the case of interest in this paper), by the presence of noise terms in the dynamical equations, by the existence of competitive interactions, heterogeneity in the network of connectivities~\cite{TessoneZanetteToral2008} or any other origin. A general theory to explain this transition has been developed in~\cite{tessone2007a}, while an exact treatment in the case of disorder in the natural frequencies has been carried out in~\cite{OttAntonsen:2009,LafuerzaColetToral2010}. In the next section we present a simple treatment of this problem in terms of the order parameter expansion, which allows one to derive equations for the macroscopic variables as a perturbative expansion, assuming small fluctuations. This simple approach is able to predict the main features observed in the numerical simulations. Furthermore, it gives access to an analytic expression for the critical noise intensities in the large coupling limit.

\section{Method}
\label{sec:method}
\subsection{Derivation of the dynamical equations}
As stated in the introduction, our goal is to use the  order parameter expansion method to obtain evolution equations for the global phase $\Psi(t)$, defined in Eq.(\ref{eq:centerMass}), and its fluctuations, defined as suitable moments of the variables $\epsilon_i(t)=\phi_i(t)-\Psi(t)$. We first notice that according to this definition and using Eq.~(\ref{eq:centerMass}), a short algebra leads to $\left\langle \rm{e}^{\imath\phi_j(t)}\right\rangle =\rm{e}^{\imath\Psi(t)}\left\langle \cos{\epsilon_j(t)}+\imath\sin{\epsilon_j(t)}\right\rangle=\rho(t){\rm e}^{\imath\Psi(t)}$. Since $\rho(t)$ has to be a real number we find that $\left\langle \sin{\epsilon_j(t)}\right\rangle =0$ and $\rho(t)=\left\langle \cos{\epsilon(t)}\right\rangle$. As a consequence we can rewrite Eq.~(\ref{eq:ActiveRots}) as:
\begin{equation}
\label{eq:ActiveRotsSymmetric}
 \dot \phi_i(t)=\omega_i-\sin(\phi_i(t))-C \left\langle \cos{\epsilon_j(t)}\right\rangle \sin{\epsilon_i(t)} \,.
\end{equation}

If we now write $\delta_i=\omega_i-\left\langle \omega\right\rangle $ for the deviation of the local natural frequency from the mean and take then the time derivative of~(\ref{eq:centerMass}) one can identify real and imaginary parts and find the identity $\dot\Psi(t)\left\langle \cos{\epsilon_j(t)}\right\rangle =\left\langle \dot\phi_j(t)\cos{\epsilon_j(t)}\right\rangle $. There we substitute $\dot\phi_j(t)$ by Eq.~(\ref{eq:ActiveRotsSymmetric}) and obtain an equation for $\dot\Psi(t)$ as a function of $\left\langle \cos{\epsilon_j}\right\rangle $, $\left\langle \cos^2{\epsilon_j}\right\rangle $, $\left\langle \sin{\epsilon_j}\cos{\epsilon_j}\right\rangle $ and $\left\langle \delta_j\cos{\epsilon_j}\right\rangle $. If we now expand these four terms up to second order around $\epsilon_i=0$ we are left with:
\begin{equation}
\label{eq:odePsi}
 \dot\Psi = \left\langle\omega\right\rangle - \Omega_2(t)\sin{\Psi(t)} \,.
\end{equation}
Here we have identified the dynamical variable $\Omega_2(t)=1-\displaystyle\frac{\left\langle \epsilon_j^2(t)\right\rangle }{2}$. We determine its dynamics by writing $\dot\Omega_2(t)=-\left\langle \dot\epsilon_j(t)\epsilon_j(t)\right\rangle $, using $\dot\epsilon_i(t)=\dot\phi_i(t)-\dot\Psi(t)$, and replacing $\dot\phi_i$ from Eq.~(\ref{eq:ActiveRotsSymmetric}) and $\dot\Psi$ from Eq.~(\ref{eq:odePsi}). Expanding the resulting expression up to second order around $\epsilon_i=0$ we arrive at:
\begin{equation}
\label{eq:odeRho}
 \dot\Omega_2 = - W(t) + 2\left( \cos{\Psi(t)}+C\right)\left(1-\Omega_2(t)\right)\,,
\end{equation}
where the third dynamical variable $W(t)=\left\langle \epsilon_j(t)\delta_j\right\rangle$ allows us to close the set of equations. It obeys dynamics given by $\dot W(t)=\left\langle \dot\epsilon_j(t)\delta_j\right\rangle $ and is found in the same way as above:
\begin{equation}
\label{eq:odeW}
 \dot W=\sigma^2-\left(\cos{\Psi(t)}+C\right)W(t)\,,
\end{equation}
where we made use of the definition $\left\langle \delta_j^2\right\rangle =\sigma^2$. 

The set of equations (\ref{eq:odePsi}) for the global phase and~(\ref{eq:odeRho}-\ref{eq:odeW}) for the fluctuations, is the result of the  order parameter expansion applied to the oscillator ensemble defined by Eqs.~(\ref{eq:ActiveRots}) and is the basis of our subsequent analysis. The errors are of the order $O\left(\left\langle \delta_j^n\epsilon_j^m\right\rangle\right), {n+m=3}$, or higher. As a consequence, Eq.~(\ref{eq:odePsi}) is more accurate than the corresponding equation $\dot\Psi=\left\langle\omega\right\rangle/\rho-\sin{\Psi}+O\left(\left\langle \delta_j^2\right\rangle \right)$ obtained in~\cite{tessone2007a}. Note that this last equation simply identifies $\rho$ as the threshold for excitability. In our case, the full stability analysis is more involved as $\Omega_2(t)$ is considered to be a variable of time. In the next subsection we will determine the fixed points of the system~(\ref{eq:odePsi}-\ref{eq:odeW}) and their stability.

\subsection{Phase diagram}
The fixed points $(\Psi^*,\Omega_2^*,W^*)$ of the system of equations ~(\ref{eq:odePsi}-\ref{eq:odeW}) must satisfy:
\begin{eqnarray}
\label{eq:fpPsi}
\left\langle\omega\right\rangle&=&\Omega_2^*\sin{\Psi^*}\,,\\
\label{eq:fpOmega} 
\Omega_2^*&=&1-\frac{\sigma^2}{2\left(\cos{\Psi^*}+C\right)^2}\,,\\
\label{eq:fpW} 
W^*&=&\frac{\sigma^2}{\cos{\Psi^*}+C}\,.
\end{eqnarray}

\begin{figure}
 \centering
 \includegraphics[width=1.0\columnwidth]{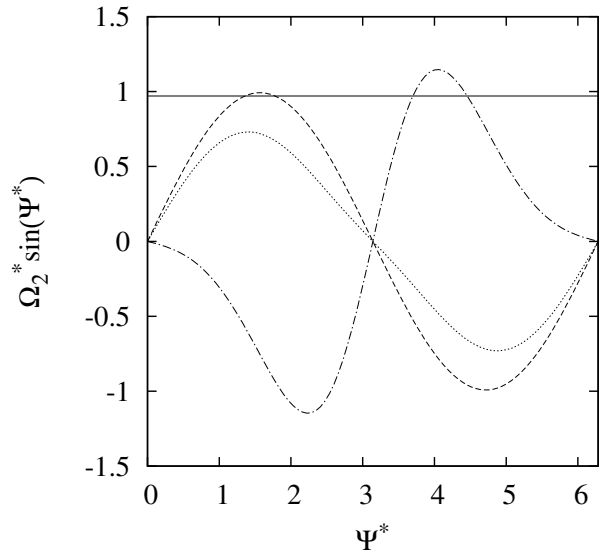}
 \caption{Graphical analysis of the solutions of the equation $\left\langle\omega\right\rangle=\left(1-\frac{\sigma^2}{2\left(\cos{\Psi^*}+C\right)^2}\right)\sin\left({\Psi^*}\right)\equiv\Omega_2^*(\Psi^*)\sin\left({\Psi^*}\right)$ for $C=4.0$ and $\sigma=0.5,3.0,7.5$ (dashed, dotted, dash dotted). The horizontal line marks $\left\langle\omega\right\rangle=0.97$. Note that the line corresponding to $\sigma=3.0$ does not cut the horizontal line, thus no stable steady state exists for this value of $\sigma$, whereas two solutions exist for the other values of $\sigma$.}
 \label{fig:nullclines}
\end{figure}

Graphically, the coordinates $\Psi^*$ of the fixed points correspond to the intersections of the function $\Omega_2^*(\Psi^*)\sin\left({\Psi^*}\right)$ with the horizontal line representing $\left\langle\omega\right\rangle$. As shown in figure~\ref{fig:nullclines}, it turns out that, for fixed $\langle \omega\rangle$ and $C$ there exist two limiting values of the diversity $\sigma_c$ and $\sigma_c'$ such that two solutions are found whenever $\sigma<\sigma_c$ or $\sigma>\sigma_c'$. A linear stability analysis shows that in this case the global phase behaves as an excitable system, corresponding one of the solutions to an stable and the other to an unstable fixed point. If, otherwise, $\sigma\in(\sigma_c,\sigma_c')$ there will be no fixed points and the global phase will rotate in time, signaling the existence of coherence firing in the global system. 
The linear stability analysis also shows that the stable and unstable fixed points, found in the low and high noise limits, collide and disappear when the maximum of the right-hand-side of Eq.~(\ref{eq:fpPsi}) coincides with $\left\langle\omega\right\rangle$. This is a so-called SNIC (saddle-node on an invariant cycle) bifurcation~\cite{Ermentrout1986}. The steady states in the macroscopic equations~(\ref{eq:odePsi}-\ref{eq:odeW}) at high and at low values of $\sigma$ are caused by different underlying microscopic dynamics: whereas individual rotators are moving at high levels of noise, they are all at rest in the low noise limit.

\begin{figure}
 \centering
\includegraphics[width=0.8\columnwidth]{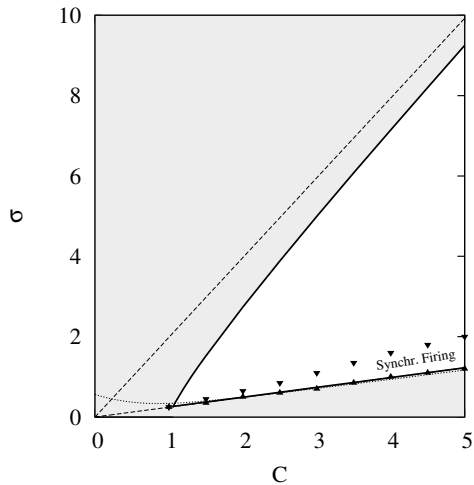}
 \caption{Phase diagram: Fixed points of Eqs.~(\ref{eq:fpPsi}-\ref{eq:fpW}) (for~$\left\langle\omega\right\rangle=0.97$) exist below and above the two continuous lines (gray region). In between no fixed points exist and the global phase rotates, i.e. the individual rotators oscillate in a coherent manner. Approximate values of the critical diversity for $C\gg 1$ according to Eqs.~(\ref{eq:PT1},\ref{eq:PT2}) are plotted with dashed lines. Symbols show values taken from numerical simulations of the set of Eqs.~(\ref{eq:ActiveRots}) with a Gaussian distribution. The dotted line marks the approximate solution of the critical diversity given in~\cite{LafuerzaColetToral2010}.}
 \label{fig:phaseDiagramm_general}
\end{figure}

The phase diagram identifying regions of synchronized global firing can be obtained from the existence of solutions to equations~(\ref{eq:fpPsi}-\ref{eq:fpW}) as discussed above. In general, this has to be performed numerically, but to an arbitrary degree of accuracy. Results for the case that the mean of natural frequencies is $\left\langle\omega\right\rangle=0.97$ are shown in figure~\ref{fig:phaseDiagramm_general}. It can be observed that a minimal coupling intensity is needed to introduce a possible state of coherent firing. In the large coupling limit, $C\gg 1$, it is possible to derive analytical expressions for $\sigma_c$ and $\sigma_c'$. Neglecting $\cos \Psi^*$ in the denominator of the right hand side of Eq.(\ref{eq:fpOmega}), the necessary condition $|\langle \omega\rangle|\le \Omega_2^*$ leads to 
\begin{eqnarray}
\label{eq:PT1}
\sigma_{c}&=&C\sqrt{2}\sqrt{1-\left\langle\omega\right\rangle},\\
\label{eq:PT2}\sigma_{c}'&=&C\sqrt{2}\sqrt{1+\left\langle\omega\right\rangle} \,.
\end{eqnarray}
In this approximation, the width of the interval $(\sigma_{c},\sigma_{c}')$, where the system fires synchronously, grows linear with $C$. This means that an intermediate level of coupling is needed to support a synchronized firing state. The dependence on $\left\langle\omega\right\rangle$ of the second transition is rather small for $\left\langle\omega\right\rangle\approx 1$. The interval collapses for $\left\langle\omega\right\rangle=0$. The resulting approximate phase diagram for large coupling values is marked with dashed lines in figure~\ref{fig:phaseDiagramm_general}. We conclude that the  order parameter expansion correctly identifies the diversity induced transitions that occur at the critical points $\sigma_c$ and $\sigma_c'$. As shown in figure \ref{fig:phaseDiagramm_general}, it also allows the determination of the value of $\sigma_c$ with a reasonable accuracy, although $\sigma_c'$ is grossly overestimated, when compared against the numerical simulations (see section~\ref{sec:numSim}) or the analytical treatment of~\cite{LafuerzaColetToral2010} using a Gaussian distribution $g(\omega_i)$ for the natural frequencies. As an example, for $\left\langle\omega\right\rangle=0.95$ and coupling $C=4$, the numerical solution of Eqs.~(\ref{eq:fpPsi}-\ref{eq:fpW}) yields a critical noise intensity of $\sigma_c=1.269$, whereas the approximate solution, Eq, (\ref{eq:PT1}) yields $\sigma_c=1.265$. This is to be compared with the value $\sigma_c=1.272$ obtained from the exact treatment given in \cite{LafuerzaColetToral2010} based on recent developments by Ott and Antonsen~\cite{OttAntonsen:2008,OttAntonsen:2009}.

\begin{figure}
 \centering
\includegraphics[width=0.8\columnwidth]{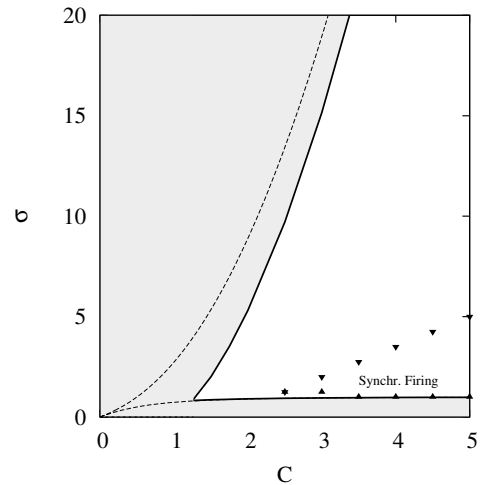}
 \caption{Phase diagram for the exponential distribution of natural frequencies which satisfies $\sigma=\left\langle\omega\right\rangle$. As in figure~\ref{fig:phaseDiagramm_general}, fixed points of Eqs.~(\ref{eq:fpPsi}-\ref{eq:fpW}) exist below and above the two continuous lines (gray region). Approximate values of the critical diversity for $C\gg 1$ according to Eqs.~(\ref{eq:PT1expo}-\ref{eq:PT2expo}) are plotted with dashed lines. Symbols show numerical simulations with exponential distribution.}
 \label{fig:phaseDiagramm_expo}
\end{figure}

From the microscopic point of view, one could argue that the destruction of coherence at high noise values is due to the coexistence of individual oscillators rotating at opposite directions, as they would certainly be present for many general distributions $g(\omega)$ of natural frequencies. However, the only requirements we have made on the distribution $g\left(\omega\right)$ is that its first and second moments are well defined. Therefore, according to our treatment, the existence of elements rotating in both directions can not be the only responsible for the transitions. To analyze this issue, we have considered that the individual frequencies were drawn from an exponential distribution $g(\omega_i)=\rm{e}^{-\omega_i/\left\langle\omega\right\rangle}/\left\langle\omega\right\rangle$, for $\omega_i\ge 0$ such that all natural frequencies $\omega_i$ would be positive. In this case the variance $\sigma^2$ and the mean $\left\langle\omega\right\rangle$ are not independent of each other, as they satisfy $\sigma=\left\langle\omega\right\rangle$ and there is only one parameter in the distribution. Replacing $\sigma=\left\langle\omega\right\rangle$ in Eqs.~(\ref{eq:PT1}) and~(\ref{eq:PT2}) we obtain
\begin{eqnarray}
\label{eq:PT1expo}
\sigma_{c}&=&C\left(\sqrt{C^2+2}-C\right)\\
\label{eq:PT2expo}
\sigma_{c}'&=&C\left(\sqrt{C^2+2}+C\right) \,,
\end{eqnarray}
as the limits of the zone for which synchronized firing exists. 
The phase diagram for this exponential distribution has been plotted in figure~\ref{fig:phaseDiagramm_expo}. As it is a special case of the general distributions considered above, the qualitative image is the same: an intermediate value for the intensity of the quenched noise is required to induce a state of coherent firing, while a too high intensity destroys it. As in the case of Gaussian distribution of natural frequencies, the qualitative picture agrees with the exact treatment and the numerical simulations. The lower critical point $\sigma_c$ is also given with a reasonably degree of accuracy, but the upper critical point is overestimated, again compared with the numerical simulations or the analytical treatment of~\cite{LafuerzaColetToral2010}.

To end this section, we note that, for a general distribution $g(\omega)$, a very large diversity satisfying $\sigma>\sigma_c'$ will never induce another SNIC bifurcation into a new state of coherent firing. With this observation one would expect that distributions $g(\omega)$ with infinite variance, as is the case for a Lorentzian distribution, would never show a regime of synchronized firing. This is in agreement with the detailed theory of ~\cite{LafuerzaColetToral2010} only for $\left\langle\omega\right\rangle<1$. In the case $\left\langle\omega\right\rangle>1$, however, the complete theory predicts that oscillators rotate coherently for low diversity and incoherently for high diversity.

In summary, and in agreement with more involved treatments of the coupled active rotator model, the  order parameter expansion scheme predicts a transition into coherent firing and out of it, induced by the exclusive presence of quenched noise. The only assumptions we made on the frequency distribution to derive the results are the existence of well defined first and second moments. In the following section we present numerical simulations of the full system, Eqs.~(\ref{eq:ActiveRots}).

\section{Numerical simulations}
\label{sec:numSim}

In the previous section we demonstrated that for very low and very high values of $\sigma$ the system~(\ref{eq:odePsi}-\ref{eq:odeW}) is in a steady state characterized by $\dot\Psi=\dot\Omega_2=\dot W=0$, whereas for intermediate values the global phase $\Psi$ is not constant. This reproduces, in a simple manner, the prediction of the existence of this intermediate level of disorder for which the system fires synchronously and shows the validity of the  order parameter expansion applied to this model. In this section, we will present results of numerical simulations of the full system of coupled equations~(\ref{eq:ActiveRots}). Our goal is to show that the transitions occurring at $\sigma_c$ and $\sigma_c'$ show all the features of true phase transitions and can be characterized, besides by the vanishing of the order parameter, by a divergence of the fluctuations. The divergence, as usual, is smeared out by finite size effects and it is possible to carry out an analysis in terms of finite size scaling with the number $N$ of rotators~\cite{cardy}. Furthermore we want to compare the macroscopic behavior of systems with symmetrically distributed natural frequencies and systems with only positive frequencies. Namely, Gaussian distributions are used in the first case and exponential distributions in the second. 

 As order parameter, $m$, quantifying the collective firing regime we have chosen the time average of the slope of the global phase $m=\overline{\dot\Psi}$. This is expected to vanish for small, $\sigma<\sigma_c$, and large $\sigma>\sigma_c'$ disorder and be non-zero in between. In the figures we plot the ensemble average $\ensemblemean{m}$,  and the normalized fluctuations $\chi=\displaystyle\frac{N}{\sigma^2}\left[\ensemblemean{m^2} -\ensemblemean{m}^2\right]$, where $\ensemblemean{\cdots}$ denotes an ensemble average over realizations of the random noise terms and initial conditions. We present separately the results for Gaussian and for exponentially distributed frequencies.

\subsection{Gaussian distributed $\omega_i$'s}
\begin{figure}
 \centering
 \includegraphics[width=1.0\columnwidth]{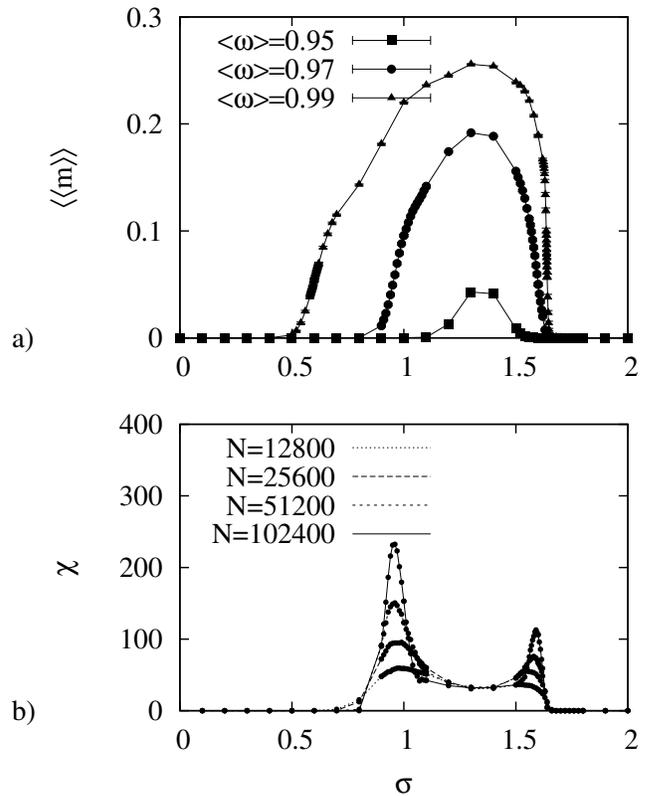}
 \caption{Simulation results for Gaussian distributed $\omega_i$'s: a)~The order parameter $\ensemblemean{m}$ for $C=4.0$ and various values of $\left\langle\omega\right\rangle$. The location of the second transition changes little by small variations in $\left\langle\omega\right\rangle$ if it is close to one. Simulations were done with $N=102400$. b) Ensemble fluctuations (performed over $1000$ realizations of the quenched noise variables and initial conditions) of the order parameter increase with system size ($\left\langle\omega\right\rangle=0.97$ and $C=4.0$).}
 \label{fig:psiDotGaussian}
\end{figure}

The natural frequencies $\omega_i$ are drawn from a Gaussian distribution of mean $\left\langle\omega\right\rangle$ and variance $\sigma^2$. In figure~\ref{fig:psiDotGaussian}a we present the results for different values of the mean frequency $\langle \omega\rangle$ as a function of the noise intensity $\sigma$. One can see that for small $\sigma$ the order parameter $\ensemblemean{m}$ vanishes or, equivalently, that the global phase is constant indicating that all oscillators are in the rest state. When reaching the critical value $\sigma_c$, the global phase $\Psi(t)$ starts to rotate, i.e. $\dot\Psi(t)\sim m\neq0$. This is the regime of synchronized firing where a macroscopic fraction of the oscillators fire in synchrony. Increasing the diversity over the second critical value $\sigma_c'$, the global phase $\Psi$ is constant again. This is the phase where all units fire in a desynchronized manner. As predicted by Eq.~(\ref{eq:PT2}) the second transition is relatively constant regarding small changes in $\left\langle\omega\right\rangle$ when it is close to one. 

The precise numerical determination of the location of the transition points $\sigma_c$ and $\sigma_c'$ is hindered by the finite size effects. We have found that the location of the maximum of the fluctuations of $m$ can give us a good estimator of the transition points, as it is relatively constant with system size, see figure~\ref{fig:psiDotGaussian}b. The results for different values of the coupling strength $C$ are indicated with symbols in the phase diagram, figure~\ref{fig:phaseDiagramm_general}. The first transition is predicted with high accuracy whereas the second transition is highly overestimated by the  order parameter expansion. Another feature predicted by the order parameter expansion, namely the existence of a minimal coupling necessary for inducing coherent firing, is indeed observed in the simulations. 

\begin{figure}
 \centering
 \includegraphics[width=1.0\columnwidth]{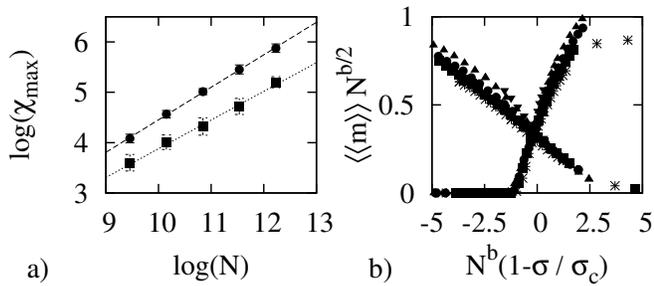}
 \caption{Finite size analysis for Gaussian distributed frequencies with $\left\langle\omega\right\rangle=0.97$ and $C=4$: a)~Linear fits of maximal fluctuations yield $\log{(\chi_{max})}\sim c\log{(N)}$ with $c=0.65\pm0.07$ for the first transition (circles) and $c=0.6\pm0.1$ for the second (squares). b)~Rescaled order parameter $\ensemblemean{m(\sigma,N)}N^{b/2}$ collapses as a function of $\epsilon N^{b}$ with exponent $b=1/3$, ($N=12800,\dots, 204800$).}
 \label{fig:scalingGauss}
\end{figure}

In the vicinity of both transitions at $\sigma_c$ or $\sigma_c'$, the ensemble fluctuations $\chi$ of the order parameter diverge with system size. As figure \ref{fig:scalingGauss}(a) shows, the maximum value $\chi_{max}(N)$ scales as $N^c$ with $c=0.65\pm0.03$ at the first transition and $c=0.61\pm0.07$ at the second. Interestingly enough, the values of the critical exponent at both transitions seem to be consistent with the value $c=2/3$ observed in a phase transition induced by quenched noise in a Ginzburg--Landau model~\cite{KT2010}. It turns out that the full dependence of $N$ and $\sigma$ at both transitions can be fitted using standard finite-size-scaling theory~\cite{cardy,deutsch:92} as $\ensemblemean{m(\sigma,N)}=N^{-b/2}f_m(\epsilon N^{b})$ and $\chi(\sigma,N)=N^{c}f_{\chi}\left(\epsilon N^{b}\right)$ with $\epsilon=1-\sigma/\sigma_c$ or $\epsilon=1-\sigma/\sigma_c'$, and being $f_m$ and $f_{\chi}$ suitable scaling functions different for each one of the transitions. Our numerical results are not sufficiently precise to allow an accurate determination of the exponent $b$, but reasonable scaling collapse of the data, see figure~\ref{fig:scalingGauss}(b), is achieved using $b=1/3$, as suggested by the analogy with the Ginzburg--Landau model mentioned before.

\subsection{Exponentially distributed $\omega_i$'s}
\begin{figure}
 \centering
 \includegraphics[width=1.0\columnwidth]{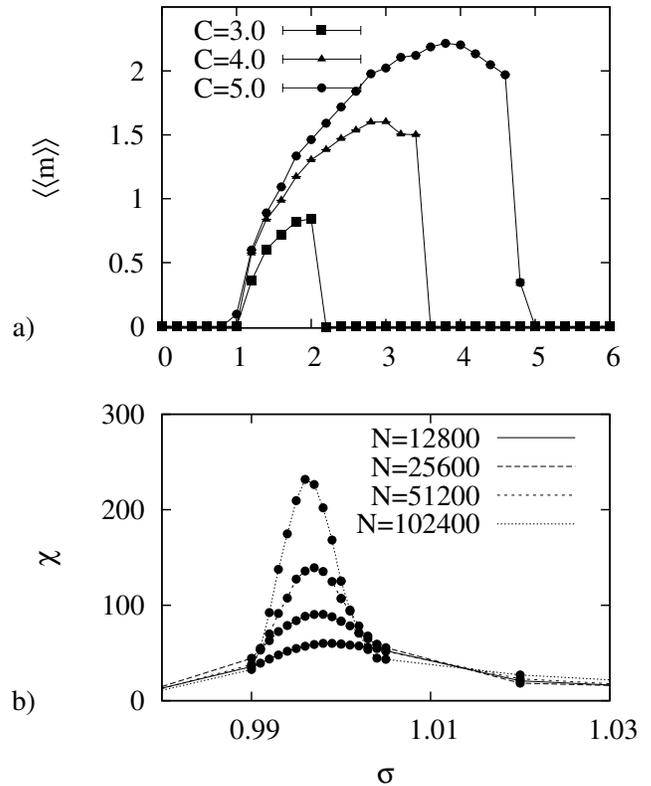}
 \caption{Simulation results for exponentially distributed $\omega_i$'s: a)~The order parameter is non-zero for finite disorder. Fluctuations show that the first transition takes place around $\sigma\approx 1$, as Eq.~(\ref{eq:PT1expo}) predicts for large $C$. b)~Ensemble fluctuations (for $C=5.0$) diverge at the first transition for increasing $N$.}
 \label{fig:psiDotExponential}
\end{figure}

The probability distribution function for the natural frequencies is $g(\omega_i)=\rm{e}^{-\omega_i/\left\langle\omega\right\rangle}/\left\langle\omega\right\rangle$ for $\omega_i\ge 0$. As mentioned before, this distribution has only one parameter as the standard deviation is equal to the mean $\sigma=\langle \omega\rangle$. It is chosen such that all rotators have natural frequencies in the same, anti-clockwise, direction. As shown in figure \ref{fig:psiDotExponential} we find the same dynamical regimes as a function of the disorder $\sigma$ than in the case of an arbitrary distribution. This is in accordance to the theoretical predictions displayed in figure~\ref{fig:phaseDiagramm_expo}. the transition into coherent firing is rather constant and happens around $\sigma_c\approx1$, the interval grows with rising coupling strength and a minimal $C$ is needed to induce coherent firing. Again the second transition is overestimated. As before, we use the maximum of the fluctuations in the order parameter (see figure~\ref{fig:psiDotExponential}b for the case of the transition at $\sigma_c$) to estimate values for the critical noise intensities and annotate them in the corresponding phase diagram (figure~\ref{fig:phaseDiagramm_expo}). 

\begin{figure}
 \centering
 \includegraphics[width=1.0\columnwidth]{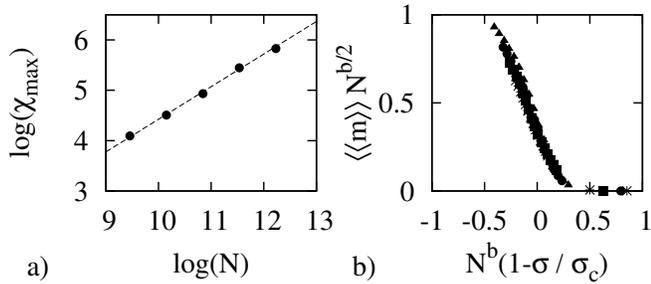}
 \caption{Finite size analysis for exponentially distributed frequencies ($C=5$): a)~Linear fit of maximal fluctuations yields $c=0.65\pm0.02$. b)~Rescaled order parameter with scaling exponent $b=1/3$.}
 \label{fig:scalingExpo}
\end{figure}

\begin{figure}
 \centering
 \includegraphics[width=1.0\columnwidth]{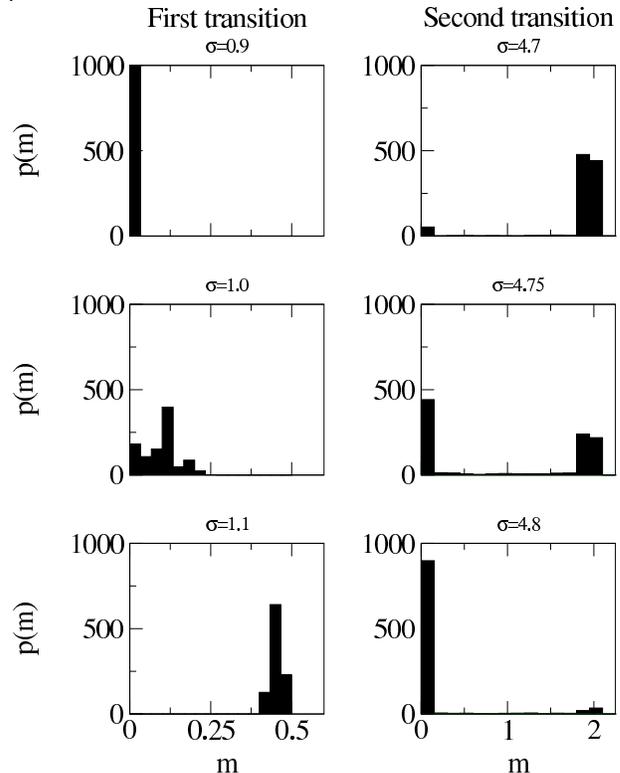}
 \caption{Histogram of 1000 steady states from simulations of equations~(\ref{eq:ActiveRots}) with exponentially distributed frequencies. At the transition into coherent firing (left column) the values are distributed around one single value, broadening near the critical disorder. The destruction of the ordered state is a first order transition (right column), the values are distributed narrowly around zero or the non-zero value, $C=5.0$.}
 \label{fig:histogramExp}
\end{figure}

The first transition, into coherent firing, is marked by diverging fluctuations (for a particular case, $C=5.0$, see figure~\ref{fig:psiDotExponential}b) which scale in the same way with system size $N$ as we have seen in the Gaussian case (see evidence in figure~\ref{fig:scalingExpo}). However, in stark contrast to Gaussian distributions, the simulations with exponential distributed frequencies give strong evidence that the transition into asynchronous firing is now of first order. We compare the histograms of steady states for 1000 noise realizations around both transitions in figure~\ref{fig:histogramExp}. At the first transition (left column) the distribution broadens at the critical disorder and moves continuously to higher values. On the contrary the equilibria near the second transition are narrowly distributed around zero, or around the non-zero value in the ordered state, typical for a first order transition (right column).

\begin{figure}
 \centering
 \includegraphics[width=1.0\columnwidth]{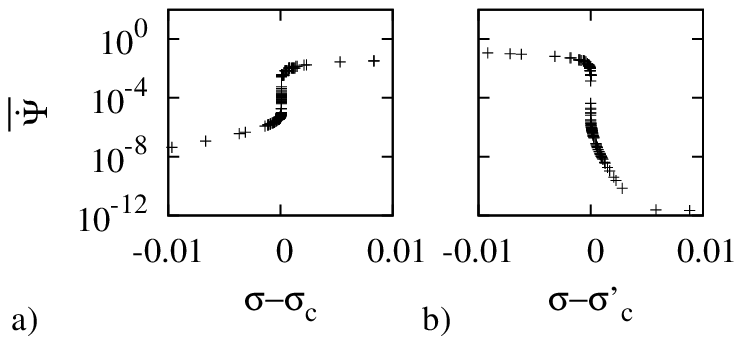}
 \caption{Results from the integration of Eqs.~(\ref{eq:odePsi}-\ref{eq:odeW}) in the vicinity of the transition points with $\left\langle\omega\right\rangle=0.97$ and $C=4.0$. Both transitions, into the ordered state and into asynchronous firing, panel a) and b) respectively, are of second order. }
 \label{fig:OPE_PTgauss}
\end{figure}

\begin{figure}
 \centering
 \includegraphics[width=1.0\columnwidth]{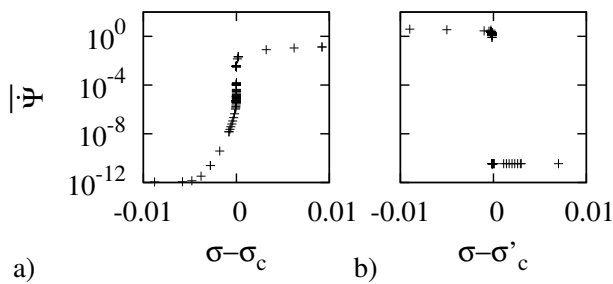}
 \caption{Results from the integration of Eqs.~(\ref{eq:odePsi}-\ref{eq:odeW}) in the vicinity of the transition points with $\left\langle\omega\right\rangle=\sigma$. Whereas the transition into the ordered state is of second order, panel a), the second transition is discontinuous, panel b).}
 \label{fig:OPE_PTexpo}
\end{figure}

It turns out that the  order parameter expansion developed in the previous section predicts that the second transition into asynchronous firing occuring at $\sigma=\sigma_c'$, is of second order for the Gaussian distribution and of first order for the exponential distribution of frequencies. The results of the numerical integration of  the system of equations~(\ref{eq:odePsi}-\ref{eq:odeW}) for selected sets of parameters ($\left\langle\omega\right\rangle$, $\sigma$, $C$) for the mean phase velocity $\overline {\dot\Psi}$ are plotted in figures \ref{fig:OPE_PTgauss} (Gaussian) and \ref{fig:OPE_PTexpo} (exponential). It is evident the jump of $\overline {\dot\Psi}$ at the second critical point $\sigma_c'$ in the case of the exponential distribution whereas it is continuous for the Gaussian distribution. The first transition to synchronized firing at $\sigma=\sigma_c$ is predicted to be continuous independently of the distribution of frequencies.

\section{Conclusions}
\label{sec:conclusions}
We have used the  order parameter expansion to approximate the dynamics of the global phase in systems of coupled active rotators under the influence of quenched disorder. The method leads straightforwardly to a system of three differential equations easier treatable than the full system and more accurate than other approximations used in previous works. In agreement with exact results for the full system, the global phase of the reduced system can undergo a transition from a rest state into a rotating regime and back into a rest state, when subjected to increasing diversity. In the rest states, $\Psi(t)$ is time independent and the angular speed is zero, whereas in the intermediate regime of coherent firing, $\Psi(t)$ increases with time and the angular speed adopts a non-zero value. Our treatment allows us to give analytic expressions for the critical disorder values in the limit of large coupling. We have seen that the first transition is predicted to a high degree of accuracy whereas the second is highly overestimated.

We have used numerical simulations to show that the ensemble fluctuations of the order parameter diverge at the transition points. The simulations with Gaussian distributed frequencies show continuous transitions, both in and out of the coherent firing state, but if the frequencies are distributed according to an exponential distribution (and therefore the mean and variance are varied simultaneously) then the destruction of the ordered state is achieved through a first order transition. The  order parameter expansion scheme predicts this distinction of the transitions. A finite-size scaling analysis of the numerical simulations data indicate that the critical exponents of the transitions are consistent with those found in the athermal Ginzburg-Landau model with additive quenched noise.

\section*{Acknowledgements}

We thank Luis F. Lafuerza and Lucas Lacasa for fruitful discussions. We acknowledge financial support from the EU NoE BioSim, LSHB-CT-2004-005137, and project FIS2007-60327 from MEC (Spain). NK is supported by a grant from the Govern Balear.


\bibliography{\referencias}



\end{document}